\documentstyle[12pt,aasms4,epsf]{article}
\received{00  1998}
\accepted{00 1998}
\centerline{Submitted to the Editor of the ApJ Letters 
 on June 30, 1998}
\slugcomment{Re-submitted on August 15, 1998}
\lefthead{Yusef-Zadeh, Roberts \& Goss}
\righthead{OH (1720 MHz) Observations}

\def\ddeg   {\hbox{$.\!\!^\circ$}}              
\def\deg    {$^{\circ}$}                        
\def\kms    {\hbox{km{\hskip0.1em}s$^{-1}$}}    
\def\dasec  {\hbox{$.\!\!^{\prime\prime}$}}     
\def\asec   {$^{\prime\prime}$}                 
\def\dasec  {\hbox{$.\!\!^{\prime\prime}$}}     
\def\dsec   {\hbox{$.\!\!^{\rm s}$}}            
\def\amin   {$^{\prime}$}                       

\def\kms    {\hbox{km{\hskip0.1em}s$^{-1}$}}    
\def\etal   {{\it et al. }}                     

\begin{document}
\title{High-resolution Observations of OH(1720 MHz) Masers\\
 Toward the Galactic Center}

\author{F. Yusef-Zadeh}
\affil{Department of Physics and Astronomy, Northwestern University, 
Evanston, Il. 60208 (zadeh@nwu.edu)}

\author{D. A. Roberts}
\affil{NCSA, 405 N. Mathews Ave, Urbana, IL 61801
 (dougr@ncsa.uiuc.edu)}

\author{W. M. Goss and  D.A. Frail}
\affil{National Radio Astronomy Observatory, P.O. Box 0, 
Socorro, New Mexico 87801 (mgoss@aoc.nrao.edu), (dfrail@aoc.nrao.edu)}

\author{A. J. Green}
\affil{University of Sydney, School of Physics, A28, Sydney, NSW 2006
Australia (agreen@physics.usyd.edu.au)}

\begin{abstract}

High-resolution VLA observations of 1720 MHz OH maser emission from
Sgr A East and the circumnuclear disk with spatial and spectral
resolutions of $\approx$ 2\dasec5 $\times$ 1\dasec3 and 0.27 \kms\
are reported. This follow-up observational study focuses on the recent
discovery of a number of such OH maser features and their intense
circularly polarized maser lines detected toward these Galactic center
sources.  The 1720 MHz maser line of OH arises from collisionally
excited gas behind a C-type shock and is an important diagnostic of
the interaction process that may occur between molecular clouds and
associated X-ray emitting shell-type supernova remnants.  The present
observations have confirmed that the observed Stokes $V$ signal is due
to Zeeman splitting and that the OH masers are angularly broadened by
the scattering medium toward the Galactic center.  The scale length of
the magnetic field fluctuations in the scattering medium toward the
Galactic center is estimated to be greater than 0.1-0.2 pc using the
correlation of the position angles of the scatter-broadened maser
spots.  In addition, the kinematics of the maser spots associated with
Sgr A East are used to place a 5 pc displacement between this extended
radio structure and the Galactic center.

\end{abstract}


\vfill\eject

\section{Introduction}

A number of recent studies have shown OH(1720MHz) masers, when
unaccompanied by the main transition lines at 1665 and 1667 MHz, are
associated with molecular clouds interacting with expanding supernova
remnants such as W28, W44, G359.1-0.5 (e.g. Frail, Goss \& Slysh 1994;
Yusef-Zadeh, Uchida, \&Roberts 1995; Claussen et al. 1997).  These
observations support a model in which H$_2$ molecules, having kinetic
temperature and density between 15-200 K and $10^3-10^5$ cm$^{-3}$,
respectively, collisionally pump OH molecules causing a population
inversion in the 1720 MHz transition of OH molecules (Elitzur 1976).
More recently, Wardle, Yusef-Zadeh, \& Geballe (1998) suggest that
water molecules produced behind a C-type shock must be dissociated by
the X-ray flux from the SNR in order to enhance the abundance of the
OH molecule.

The initial observations which reported the discovery of OH(1720MHz)
maser lines toward Sgr A were carried out with a spectral resolution
0.27 \kms\ and a synthesized beam of approximately 15\asec
(Yusef-Zadeh \etal 1996, hereafter YRGFG).  Seven OH(1720) masers were
found to be associated with the nonthermal Sgr A East shell at a
velocity of $\approx$50 \kms, which is close to the systemic velocity
of the adjacent +50 \kms cloud to the SNR. 
 The eighth maser spot, however, showed a
number of velocity components near 134 \kms\ at the location where the
Northern Arm of Sgr A West crosses the circumnuclear disk (CND).  In
addition, the patterns of the Stokes $V$ spectrum associated with the
Sgr A East and the CND OH(1720MHz) maser spots were opposite to each
other implying an opposite sign of the inferred magnetic field.
  We considered the OH(1720MHz) masers with two different
kinematic and polarization characteristics in Sgr A are the result of
two different events, namely the expansion of the nonthermal shell of
Sgr A East inside the 50 \kms\ molecular cloud and cloud-cloud
collisions with a relative velocity of 134 \kms\ in the CND (YRGFG).

Given the relatively low angular resolution of the previous observations,
none of the sources were spatially resolved and it remained
possible that the Zeeman effect could have been mimicked by two
overlapping maser features with slightly different velocities and
different polarization properties.  The present observations with 
an order of magnitude more favorable spatial resolution
compared to the earlier measurements show that in most cases  
the previous spectral features
are heavily scatter-broadened and are indeed spatially separated from each
other.  In addition, the magnetic field strengths determined from
Zeeman splitting of the maser spots are consistent with previous
measurements.

\section{Observations}

The A configuration of the Very Large Array of the National Radio
Astronomy Observatory\footnote{The National Radio Astronomy
Observatory is a facility of the National Science Foundation, operated
under a cooperative agreement by Associated Universities, Inc.} on
November 22, 1996 was used to observe three fields centered on the B
source ($\alpha$, $\delta$[ 1950 ] = $17^{\rm h} 42^{\rm m}$ 29\dsec96,
$-28^\circ 58^\prime 35^{\prime\prime}$ at $V_{\rm LSR}$=134 \kms),
the C source ($17^{\rm h} 42^{\rm m}$ 28\dsec07, $-28^\circ 58^\prime$
32\dasec3 at $V_{\rm LSR}$=56 \kms), and the F source ($17^{\rm h}
42^{\rm m}$ 32\dsec6, $-29^\circ 00^\prime$ 23\dasec3 at $V_{\rm
LSR}$=56 \kms) in the 1720 MHz hyperfine transition of OH molecule
(YRGFC).  Using both the right ($RCP$) and left ($LCP$) hands of
circular polarization, the 195.3 kHz bandwidth and 127 channels gave a
velocity resolution of 0.27 \kms\ after online Hanning smoothing.  The
final images had a synthesized beam of 2\dasec48 $\times$ 1\dasec28
(PA=3\ddeg9) using the robustness parameter of 1 in IMAGR of AIPS with
the exception of source B where using a robustness parameter of --5 
produces a
spatial resolution of 1\dasec6 $\times$ 0\dasec82 (PA=-6\ddeg6).  The
robustness parameter is used to apply different weighting to the
$\it{uv}$ data.  Standard calibration was carried out using 1328+307
(flux density and bandpass) and 1748-253 (complex gain and bandpass).  The
brightest isolated maser source A was used effectively to
self-calibrate all the channels before the final cubes were
constructed.  The rms noise per channel is $\approx$ 20 and 37 mJy
beam$^{-1}$ for the low-resolution and high-resolution images,
respectively.  The analysis of the $V$ signal due to the Zeeman effect
followed the least-squares procedure of Roberts \etal\ (1993) 
in order to determine the line of sight magnetic field.
The Stokes
$I$ and $V$ spectra from the brightest maser A are  shown in Figure 1.
The solid line superposed on the $V$ spectra is the derivative of the
$I$ spectra scaled by the derived magnetic field.  The
size of the sources were measured using the Gaussian-fitting algorithm
JMFIT in AIPS on the channel that showed the peak emission. 
The errors   were  determined  from the maximum and minimum 
values of the fit parameters that JMFIT gives. If these error 
parameters were zero, the errors could not be trusted and we did not
incoprporated them in our analysis.   The flux 
density,
LSR velocity, and velocity width were determined from Gaussian fits of
the spectra using PROFIT in the Groningen Image Processing System
(GIPSY).  Table 1 lists the parameters of the fits to source B 
(v $\approx$ 134 \kms) with a
synthesized beam of 1\dasec6 $\times$ 0\dasec82, while Table 2
tabulates the parameters of the sources (v $\approx$43--66 \kms) 
observed with a synthesized
beam of 2\dasec48 $\times$ 1\dasec28.  We have used the nomenclature
introduced by YRGFG.

\section{Results}  

\subsection{Sgr A West OH(1720MHz) Maser}

The +134 \kms\ velocity feature (B source in YRGFG) is spatially
extended and shows a complex velocity structure resolved into four
velocity components B1, B2, B3 and B4. 
The brightest sources B1 and B2 are partially blended spectrally but
the positions of the peak emission are spatially coincident.
Source B4 is displaced $\approx 0.4''$ from B1 and B2 but
shows a velocity similar to B1.  The highest velocity feature at 136.3
\kms\ (source B3) is displaced to the SW by about 2\asec\ from
the lower velocity components B1 and B2 at 131.8 and 133.5
\kms\,  corresponding to a velocity gradient 
$\approx$60 \kms pc$^{-1}$.  This spatial
distribution follows the N-S orientation of the ionized gas
and the tongue of neutral cloud along the extension of the Northern 
Arm of Sgr A West (e.g. Jackson et al. 1993).

Sources B1 and B2 are clearly resolved with a size of
0\dasec86$\pm0.07\times$ 0\dasec34$\pm0.24$ with PA=177\deg $\pm
6^\circ$, and 0\dasec63$\pm0.24 \times$ 0\dasec44$\pm0.17$
 with PA=168 $\pm
43^\circ$, respectively. 
The axial ratio (minor/major) values of $\approx$0.4 
and the position angles of all three sources in Table 1 are consistent
to within the errors.  The peak flux density of the B
components is reduced by about 35\% when compared with previous
measurements using a more compact configurations of the VLA (YRGFG).
It is quite possible that the observed change is due to a
low surface brightness extended component resolved out by the 
present observation.
Time variability over a ten month period is also a possibility 
that could contribute to the significant flux density variations. 
However, variability seems unlikely since all the 
components in Table 1 have lower intensities than the YRGFG results.
The 
$V$ spectrum of source B indicates a magnetic field
strength  of 4.81$\pm0.73$ mG.  This value 
is the largest magnetic field 
yet estimated in the Galactic center region from OH Zeeman
 data.  Because of the relatively
large Zeeman shift of the line profiles observed with respect to their
Doppler linewidths, Elitzur (1998) has recently argued that the
Galactic center OH(1720MHz) maser lines reported by YRGFG are saturated.  We
have used his novel approach to determine if the present data show any
signatures of maser saturation.  Sources A, C, D and E all show
signatures of saturated masers with the value of the quantity $\tau_0$
ranging between 28 and 81 (see equation 32 of Elitzur 1998 where 
 $\tau_0$
is defined as T$_b$ = T$_x$ e$^{\tau_0}$ and  T$_b$ and T$_x$ are the
maser brightness and excitation temperatures, respectively.

\subsection{Sgr A East OH(1720MHz) Masers}

The  seven sources 
detected  by YRGFG are summarized in Table 2. Two additional 
components (H \& I) are detected in the A array observations. 
The
brightness temperature of the strongest source A (size 
0\dasec74 $\pm$ 0\dasec025 $\times$ 0\dasec55 $\pm$ 0\dasec025) is
$\approx 8.1\times10^5$ K implying that the OH (1720MHz) emission is
produced under non-thermal conditions.  The axial ratios of the sources
with the most favorable signal to noises 
(A, D, E and G1 in Table 2) 
are in the range 0.72 to 0.81. 
The position angle of their semi-major axis vary
significantly between sources A and the remainder of the resolved sources.
We also note that the sources surrounding the non-thermal shell of Sgr
A East, as discussed below, show two distinct velocity components. The
maser features located to the SW of the shell have velocities
ranging between 53 and 66 \kms\, whereas source C to the NW of
the shell has a velocity of 43 \kms.

\section{Discussion}

\subsection{The Nature of Scattering}

Recent observations show that the interstellar broadening of Sgr A$^*$
and of OH/IR stars within the inner 45\amin\ of the Galactic
center is  anisotropic (Lo \etal\ 1993; van Langevelde 
\etal\ 1992; Yusef-Zadeh \etal\ 1994; Frail \etal 1994; Lazio and
Cordes 1998).  The scatter-broadened image of Sgr A$^*$ is elongated
in the East-West direction, with an axial ratio of 0.60$\pm$0.05 at 
$\lambda$20 cm and a
position angle of 87$^0\pm$3$^0$.  Both the major and minor axes
follow the $\lambda^2$ law appropriate for scattering by turbulence in
the intervening medium.  The size of the heavily anisotropically scattered
OH/IR stars at 1612 MHz based on A-array observations ranges between
0\dasec5 and 1\dasec5 (Frail et al. 1994).
  The anisotropy is
considered to be caused 
by a magnetic field permeating the scattering medium
(e.g. van Langevelde et al. 1992), and that
the scattering is agued to occur
 within extended HII regions lying in the central
100 pc of the Galaxy (Yusef-Zadeh et al. 1994). 

Assuming that sources A to I are single sources and that their sizes
are not intrinsic, the measured  sizes agree very well with the heavily
scattered radio OH/IR stars in the Galactic center region
(Frail et al. 1994). Source C, however, is unresolved suggesting that 
the scattering medium is inhomogeneous. 
 The sources listed in Tables 1 and 2 also show a
 degree of anisotropy with the Sgr A West B sources having typical
axial ratio of $\approx$0.4 but somewhat less than 
the values of $\approx$0.7 for the 
Sgr A East maser sources. It is known that Sgr A East
lies behind Sgr A West (Yusef-Zadeh and Morris 1987; Pedlar et
al. 1989). Thus the larger degree of anisotropy for the B sources may indicate 
averaging of the sources over a larger pathlength (see below). 
We also note a trend in the position angle distribution of
the scattered sources. The clusters of resolved sources in close 
proximity appear to 
show similar position angles,  
whereas the sources that are more displaced 
show no
 apparent correlation in their position angles.  Figure 2 shows a $\lambda$6cm
continuum image of Sgr A East and Sgr A West with a resolution of
0.67$''\times0.40''$ (PA=7$^0$).  The shapes of the resolved sources in
Tables 1 and 2 are drawn on Figure 2 (the actual sizes are 15 times smaller
than those drawn).
Two other heavily scattered sources are also drawn, namely
Sgr A$^*$ and OH359.986--0.061 (1612 MHz) as discussed below.

The variation of the position angle of  scatter broadened radio
sources can be used as a powerful probe of the structure of the
ionized magnetized medium toward the Galactic center.  Using Sgr A$^*$
with a position angle of 82$^\circ \pm$ 1\ddeg8 at $\lambda=$20.7cm
(Yusef-Zadeh et al. 1996) and the Sgr A West sources (sources B1, B2
and B4 in Table 1) with similar position angles 
(in the range 177 and 23 degrees, see Figure 2) we can examine the spatial
scale of the magnetic fluctuations.  The position angles of the B
sources and Sgr A$^*$, separated by about 45\asec\ (
$\approx$1.8 pc at the distance of 8.5 kpc), show no correlation 
whereas the B sources that
are separated  by $\approx3''$ show a reasonable correlation.  This implies
that the thickness of the magnetoionized screen must be between
3$''$ and 45$''$.  If the thickness of the screen, assumed to
be near the Galactic center,  is much larger than the spatial
scale 0.1 pc, averaging along the line of sight would substantially
reduce the correlation of the orientation of the scattered
sources. This picture is also consistent with the lack of correlation
between the position angles of Sgr A$^*$ and the  nearest OH/IR star
(OH359.986-0.061) is shown as the ellipse to the NE in Figure 2
(Frail et al. 1994). This star lies about 160$''$  from Sgr A$^*$
corresponding to a projected linear separation of $\approx$ 6 pc,
 much larger
than the estimated scale length of the magnetic fluctuations.
    
The above interpretation based on limited number of 
scattered sources can also be applied to the OH(1720MHz) masers
associated with Sgr A East. While sources D and E, separated by 
$\approx$4$''$  show position angles that agree with each
other (29$^\circ, 48^\circ$) within the errors, source A and to some
extent source G1 show discrepant position angles of 147$^\circ$ and
4.5$^\circ$, respectively.  Sources A and G1 are displaced by more
than 20$''$ (0.8pc) from sources D and E. The lack of correlation
suggests a scale length of magnetic fluctuations greater than 0.1 pc
(the distance between sources D and E) and less than 0.8 pc.

It is evident from the above analysis that the power in the smallest
length scale in the turbulent medium with a thickness of greater than
about 0.1 pc varies considerably.  Recent analysis of the rotation
measure structure function of the Faraday screen toward the non-thermal
filamentary structure G359.54+0.18 suggests a thickness which is about
0.1 pc (Yusef-Zadeh, Wardle, \& Parastaran 1997). In this picture the
variation of the rotation measure on scales of about 0.1 pc was used
to  estimate  the thickness of the magnetoionized
medium. With the limited number of maser sources available as a
function of angular separations, it is striking  that the 
scale length of the magnetic fluctuations for the regions of 
Sgr A and G359.5+0.18  based on two 
different measurements are similar.

\subsection{Tidal Shear in Sgr A East}

The maser sources arise along the edge of the Sgr A East shell 
where
the acceleration is considered to be 
perpendicular to the line of sight and where the velocity coherence 
is achieved with a small velocity gradient along the line of sight. This
picture is consistent with the morphology of NH$_3$ and CS molecular 
features
thought to be the result of the compression of the 50 \kms\ molecular
cloud as the shock is driven into the cloud (Ho et al. 1991; Serabyn et al.
 1992).  The maser sources
detected to the SE of Sgr A East lie close to the elongated
compressed CS feature.  Earlier measurements of Galactic SNR's
interacting with molecular clouds showed a similarity between radial
velocities of the OH(1720MHz) masers distributed at the edge of the
remnants and the  systemic velocities of the disturbed molecular
clouds (Frail et al. 1996; Green et al. 1997). It is plausible that the large velocity difference of masers
seen in Sgr A East may be the result of a strong tidal sheer that the
50 \kms\ molecular cloud is experiencing.  The velocity gradient of
maser sources across the SE and NW edges of the Sgr A East shell is
estimated to be $\approx$2 \kms\ pc$^{-1}$.  The average 
velocity difference between
sources C and A-G is $\Delta V$ = 12 \kms\ with  a linear separation of
$\approx$6 pc.  Assuming a mass distribution $M(r) \propto r^{1.2}$ 
(e.g. Morris and Serabyn 1997; Mezger 1997), 
the above parameters suggest that Sgr A East has to 
be located to within 5 pc behind the
Galactic center (Yusef-Zadeh and Morris 1987; Pedlar et
al. 1989). 

In summary, we have reported high resolution observations of 
OH(1720) MHz masers associated with a nonthermal radio source (Sgr A East)
and the thermal source (Sgr A West), both of which are interacting with 
their corresponding molecular clouds. 
The study of the OH(1720MHz) masers in the Galactic center 
region have provided us with important 
physical quantities such as  the line of sight magnetic field behind 
a shock front, the size and shape of scatter-broadened sources 
masked by the turbulent screen toward the Galactic center,
the length scale of the magnetic fluctuations in the scattering medium
and the magnitude of the tidal shear experienced by   an expanding 
nonthermal shell into a dense molecular cloud.

\acknowledgments

F. Yusef-Zadeh's work was supported in part by NASA grant NAGW-2518.
D. Roberts acknowledges support from  NSF grant AST94-19227. We 
thank Mark Wardle and Mark Claussen for useful discussion.

\begin{figure}
\figcaption{ The Stokes $I$ and $V$ spectra from the most intense  maser A in 
Sgr A East.
The solid line superposed on the $V$ spectra is
the derivative of the respective $I$ spectra scaled by the derived
line of sight magnetic field (3.68$\pm$0.12) due to Zeeman splitting.  
}
\end{figure}

\begin{figure}
\figcaption{A $\lambda$6cm continuum image of the shell-like 
nonthermal source Sgr
A East and a spiral-like Sgr A West (saturated) 
 with a resolution of 0.67$''\times0.40''$ (PA=7$^0$).
The shapes  of the resolved sources in Tables 1 and 2
are drawn but the sizes are drawn 15 times bigger than the actual values. 
Two other heavily scattered sources are also indicated, 
Sgr A$^*$ close to the center of Sgr A West 
and OH359.986--0.061 to  the northwest of the Sgr A East shell.}

\end{figure}

\begin{references}

\reference{c97} Claussen, M.J., Frail, D.A., Goss, W.M., \& Gaume,
R.A. 1997, ApJ 489, 143.

\reference{e76} Elitzur, M. 1976, ApJ, 203, 124

\reference{e96} Elitzur, M. 1998, ApJ, in press. 

\reference{f94} Frail, D.A., Diamond, P.J., Cordes, J.M. and van
  Langevelde, H.J. 1994, ApJ, 427, L43 

\reference{f96} Frail, D. A., Goss, W. M., Reynoso, E. M., Green,
  A. J. \& Otrupcek, R. 1996, AJ, 111, 1651 

\reference{fgs94} Frail, D.A., Goss, M.W. and Slysh, V.I. 1994, ApJ,
  424, L111
\reference{g114} Green A.J., Frail, D.A., Goss, W.M. \& Otrupcek, R. 
 1997, AJ, 114, 2058. 


\reference{h93} Ho, P.T.P., Ho, L.C., Szczepaniski, J.C. Jackson, J.M.,
 Armstrong, J.T. \&
Barrett, A.H. 1991, Nature, 350, 309.



\reference{j93} Jackson, J.M., Geis, N., Genzel, R., Harris, A.I.,
  Madden, S., Poglitsch, A., Stacey, G.J., Townes, C.H. 1993, ApJ 402,
  173.


\reference{l98}  Lazio, T.J. \& Cordes, J.   ApJS, in press.
 

\reference{l93} Lo,  K.Y., Backer, D.C., Kellermann, K.I., Reid, M., Zhao,
  J.H. \etal, 1993, Nature, 362, 38.

\reference{m97} Mezger, P.G. 1997, IAUS 184, 168.

\reference{ms97} Morris, M. \& Serabyn, G. 1997, ARA\&A 34, 645.


\reference{p89} Pedlar, A., Anantharamaiah, K.R., Ekers, R.D., Goss, W.M.,
  van Gorkom, J.H., Schwarz, U.J., \& Zhao, J.-H. 1989, ApJ 342, 769.

\reference{r93}  Roberts, D.A., Goss, W.M. 1993, ApJS, 86, 133
\reference{s92} Serabyn, E., Lacy, J.H., \& Achtermann, J.M. 1992, 
ApJ, 395, 166.  

\reference{w98} van Langevelde, H.J., Frail,  
D.A., Cordes, J.M., \& Diamond, P.J. 1992, ApJ, 396, 686
 

\reference{w98} Wardle, M., Yusef-Zadeh, F., \& Geballe 1998, submitted to ApJ.

\reference{y94} Yusef-Zadeh, F., Cotton, W., Wardle, M., Melia, F. \&
  Roberts, D.A. 1994, ApJ, 434, L63.

\reference{y95} Yusef-Zadeh, F., Uchida, K.I., \& Roberts, D.A. 1995,
  Science 270, 1801. 

\reference{y96} Yusef-Zadeh, F., Roberts, D.A., Goss, W.M., Frail, D. \&
  Green, A. 1996, ApJ 466, L25 (YRGFG).

\reference{y97} Yusef-Zadeh, F., Wardle, M., \& Parastaran, P. 1997,
  ApJ 475, L119.

\reference{ym97} Yusef-Zadeh, F. \& Morris, M. 1987,
ApJ, 320, 545.

\end{references}
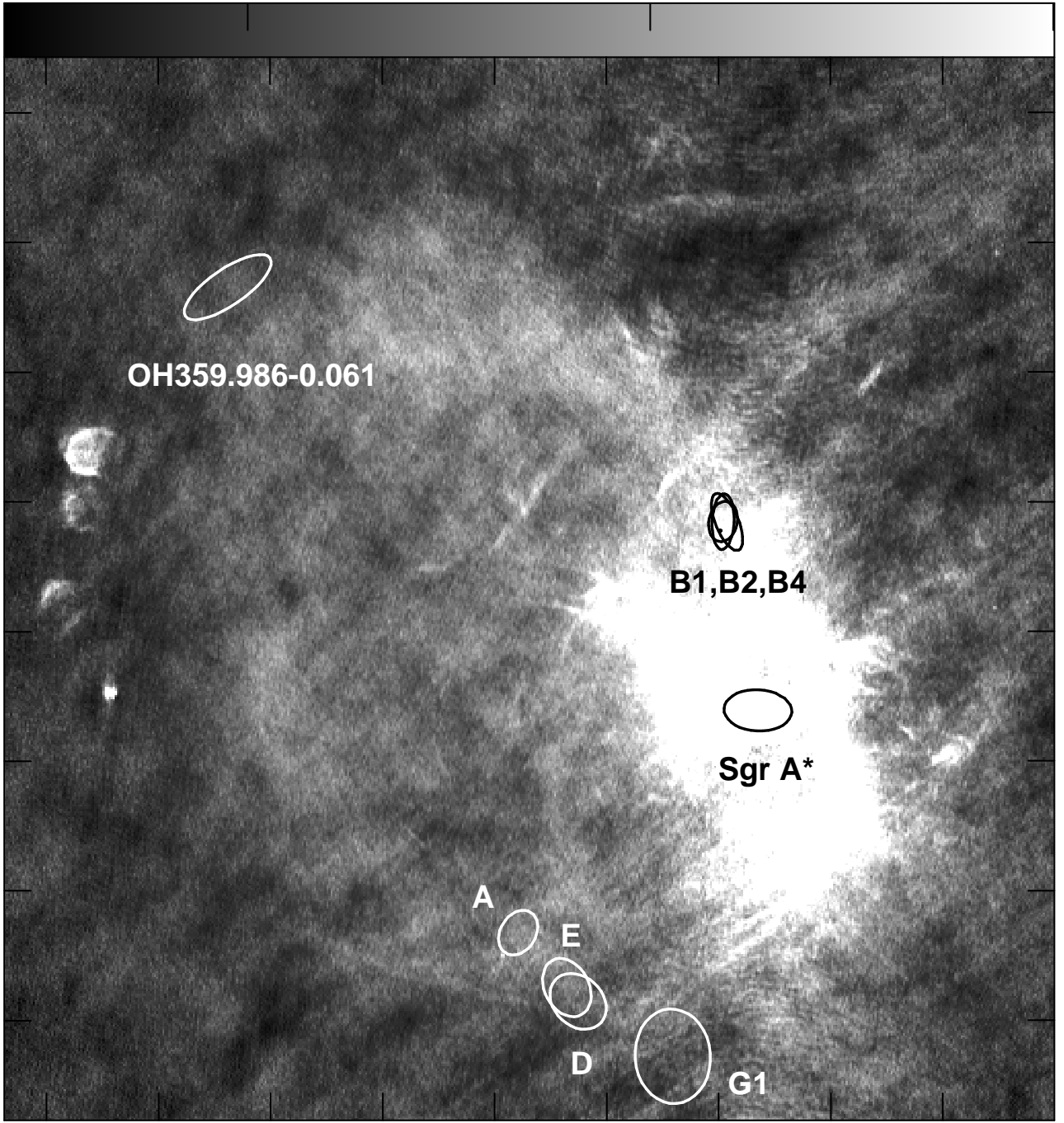
\end{document}